\documentclass[preprint,12pt]{elsarticle}

\usepackage{amssymb}
\usepackage{mathtools}
\usepackage{amsmath}
\DeclareMathAlphabet{\mathpzc}{OT1}{pzc}{m}{it}

\usepackage{siunitx}
\usepackage[hidelinks]{hyperref}
\usepackage{cleveref}
\usepackage{xcolor}
\usepackage{xargs}
\usepackage{soul}
\usepackage{subcaption}
\usepackage{nicefrac}
\usepackage{macros}
\usepackage[super]{nth}

\usepackage[mode=buildnew,subpreambles=true]{standalone}
\usepackage{shellesc}
\usepackage{tikz,tikzscale}
\usetikzlibrary{arrows,calc,fit,patterns,positioning,spy,backgrounds,decorations.fractals,shapes.misc}
\tikzset{boximg/.style={remember picture,black,thick,draw,inner sep=0pt,outer sep=0pt}}
\usetikzlibrary{external}
\usepackage{pgf}
\usepackage{pgfplots}
\usepgfplotslibrary{statistics}
\usepgfplotslibrary{fillbetween}
\usepgfplotslibrary{colormaps}
\pgfplotsset{compat=1.16}
\pgfplotsset{lua backend=true}

\journal{Computer and Fluids}

\begin{document}

\begin{frontmatter}

\title{Comparison of Entropy Stable Collocation High-Order DG Methods for Compressible Turbulent Flows}

\author[label1]{Anna Schwarz\corref{cor}\fnref{fn}} %
\author[label1]{Daniel Kempf\corref{cor}\fnref{fn}} %
\author[label1]{Jens Keim} %
\author[label1]{Patrick Kopper} %
\author[label2]{Christian Rohde} %
\author[label1]{Andrea Beck} %

\cortext[cor]{Corresponding author}
\fntext[fn]{D. Kempf and A. Schwarz share first authorship.}
\affiliation[label1]{organization={Institute of Aerodynamics and Gas Dynamics, University of Stuttgart},%
            addressline={Wankelstr. 3},
            city={Stuttgart},
            postcode={70563},
            country={Germany}}
\affiliation[label2]{organization={Institute of Applied Analysis and Numerical Simulation, University of Stuttgart},%
            addressline={Pfaffenwaldring 57},
            city={Stuttgart},
            postcode={70569},
            country={Germany}}

\begin{abstract}
High-order methods are well-suited for the numerical simulation of complex compressible turbulent flows, but require additional
stabilization techniques to capture instabilities arising from the underlying non-linear hyperbolic equations.
This paper provides a detailed comparison of the effectiveness of entropy stable discontinuous Galerkin methods for the
stabilization of compressible (wall-bounded) turbulent flows.
For this investigation, an entropy stable discontinuous Galerkin spectral element method is applied on Gauss--Legendre and Gauss--Lobatto nodes.
In the compressible regime, an additional stabilization technique for shock capturing scheme based on a convex blending of a low-order finite volume with the high-order discontinuous Galerkin operator is utilized.
The present investigation provides a systematic study from convergence tests, to the Taylor--Green vortex and finally to a more
intricate turbulent wall-bounded 3D diffuser flow, encompassing both weakly compressible and compressible regimes.
The comparison demonstrates that the DGSEM on Gauss--Lobatto nodes is less accurate due to the lower integration accuracy.
Conversely, it is faster than the DGSEM on Gauss–Legendre nodes due to a less severe time step restriction and simpler numerical operator.
To the author's knowledge, this is the first time for which a comparison of entropy stable DGSEM on Gauss--Lobatto and
Gauss--Legendre has been performed for compressible, wall-bounded turbulent flows with separation.
\end{abstract}


\end{frontmatter}

\section{Introduction}
\label{sec:introduction}

High-order methods are promising candidates for the simulation of complex multi-scale flows due to their favorable dispersion and
dissipation properties compared to low-order schemes~\cite{Ainsworth2004a,Gassner2012}.
Among the high-order methods, the discontinuous Galerkin (DG) scheme and related methods are particularly suitable for the simulation of complex,
time-dependent problems on unstructured meshes~\cite{Beck2014,Beck2018}. DGSEM is a special class of DG methods which approximates the nodal solution by a tensor-product of 1D Lagrange polynomials and collocates the interpolation and quadrature nodes.
Despite their favorable properties, high-order methods are subject to instabilities for non-linear hyperbolic equations such as
aliasing, while low-order methods are commonly equipped with an intrinsic amount of dissipation to stabilize the flow.
As such, high-order methods are generally combined with additional stabilization techniques such as artificial viscosity~\cite{Persson2006}, slope limiting~\cite{Krivodonova2007a}, filtering~\cite{Hesthaven2008}, polynomial de-aliasing through over-integration~\cite{Kirby2003} or entropy stable schemes~\cite{Fisher2013,Carpenter2014}.

The first entropy conserving numerical methods for non-linear hyperbolic conservation laws were developed
by~\citet{Tadmor1987}. The author proposed an entropy conservative low-order finite volume scheme by relying on an appropriate
entropy conservative flux that fulfills Tadmor's discrete entropy condition. This method was extended to high-order finite
difference schemes on structured grids in~\cite{LeFloch2002}, unstructured grids in~\cite{Ray2016} and to high-order finite volume methods
(ENO) in~\cite{Fjordholm2012}.
Fisher and Carpenter~\cite{Fisher2013,Carpenter2014} extended these entropy stable schemes to high-order operators on unstructured
meshes for the Euler and Navier--Stokes equations on finite domains. Their idea relies on a diagonal mass matrix, the summation-by-parts (SBP)
property, and the flux differencing form. The latter allows to write the derivative of the non-linear flux similar to a low-order FV scheme
such that a two-point flux which satisfies Tadmor's discrete entropy condition~\cite{Tadmor1987} can be employed.
This idea was extended to construct high-order entropy stable schemes for the ideal magneto-hydrodynamics equations or shallow water
equations~\cite{Gassner2016a,Winters2017}, or to unstructured triangular meshes~\cite{Crean2017}.
It is also possible to construct kinetic energy preserving schemes by relying on split forms of conservation laws, given a diagonal-norm SBP operator~\cite{Gassner2016}.
This applies also to the construction of entropy conservative or entropy stable schemes on unstructured grids for the compressible Euler or Navier--Stokes
equations by relying on diagonal norm SBP operators which are straight-forward to construct for tensor-product
elements~\cite{Gassner2014,Ortleb2017}.
Recently, it was shown that the diagonal norm SBP operator on Gauss--Legendre nodes can also be written in a flux differencing
form~\cite{Mateo-Gabin2023} which is equivalent to the original version.

An additional source of instabilities arising from the discretization of non-linear hyperbolic conservation laws with high-order methods are so-called Gibbs oscillations which are visible in the vicinity of physical discontinuities such as shocks.
Shock capturing procedures can be used to tackle these oscillations.
In the context of high-order methods, shock capturing can be split into two steps: First, the detection of troubled-cells, i.e., cells in
which the solution is heavily oscillating due to discontinuities.
This can be tackled by analytical indicators, e.g.,~\cite{Jameson1981,Ducros1999,Persson2006,Sonntag2017}, or guided by machine learning~\cite{Ray2019,Beck2020,Schwarz2022a}.
The second step focuses on the robust but accurate numerical
treatment of discontinuities which can be achieved by adding additional numerical viscosity to the flow solution near the
gradient through, e.g., a locally $h$-refined
low-order finite volume (FV) scheme~\cite{Sonntag2017}, artificial viscosity~\cite{Persson2006}, flux correction \cite{Kuzmin2005},
or a convex blending of a high-order with a lower-order operator~\cite{Hennemann2021}.

So far, a comprehensive investigation of the performance of entropy stable DGSE schemes on Gauss--Legendre or Gauss--Lobatto nodes for
complex compressible turbulent flows remains absent.
The primary objective of this paper is to address this discrepancy by offering a systematic comparison of the performance of entropy
stable DGSE schemes on Gauss--Legendre~\cite{Chan2018,Mateo-Gabin2023} and Gauss--Lobatto nodes~\cite{Gassner2016,Ranocha2021} for
turbulent flows.
The performance of these schemes is evaluated in terms of accuracy, stability, and efficiency.
The paper begins with an easily reproducible manufacturing solution and progresses to more intricate applications, including the Taylor–-Green vortex and wall-bounded turbulent flow in a 3D diffuser, in both the weakly compressible and compressible regimes.
The shock capturing procedure relies on ideas from the convex blending approach proposed by~\cite{Hennemann2021}.
This paper will focus on the comparison of the entropy stable DGSEM on Gauss--Lobatto or Gauss--Legendre nodes and will
not detail the influence of split and/or numerical fluxes, different shock capturing procedures, or the entropy projection
required for the entropy stable Gauss--Legendre DGSEM on the performance of both approaches.
A plethora of publications have provided comprehensive comparisons of the performance of different spatial discretizations or
high-order methods for turbulent flows~\cite{Zingg20006,Larsson2023,Chapelier2024}, only to mention some.
\citet{Ortleb2017} compared the performance of kinetic energy preserving 2D DG schemes on Gauss--Lobatto or Gauss--Legendre nodes and demonstrated that Gauss--Legendre nodes show superior accuracy for viscous compressible flows.
Detailed investigations on the influence of split and numerical fluxes on the accuracy and stability have been already performed
for entropy stable and/or kinetic energy preserving DGSEM on Gauss--Lobatto nodes~\cite{Flad2017,Hindenlang2019} and also for
turbulent flows~\cite{Winters2018a}.
Chan et al.~\cite{Chan2018,Chan2022a} discussed the impact of the entropy projection required in DGSEM on Gauss--Legendre nodes on the efficiency and accuracy of the solution for under-resolved flows.
To the authors' knowledge, this is the first time that such a detailed study has been carried out to compare the performance of entropy stable DGSE schemes on Gauss--Legendre or Gauss--Lobatto nodes for complex compressible turbulent flows.

The outline of this paper is as follows:
First, briefly the governing equations and entropy variables are reviewed in~\cref{sec:theory}.
Subsequently, the entropy stable DGSEM on unstructured hexahedral meshes is presented in~\cref{sec:methods}.
In~\cref{sec:application}, the implementation is verified and various applications starting from an easily reproducible
manufacturing solution and progressing to more intricate applications, including the Taylor–-Green vortex and the
wall-bounded, separated turbulent flow in a 3D diffuser, followed by a brief conclusion in~\cref{sec:conclusion}.

\section{Fundamentals}
\label{sec:theory}
\subsection{Governing equations}
\label{sec:theory:nse}
In this work, a compressible, viscous fluid is considered which is governed by the Navier--Stokes equations (NSE), given as
\begin{align}
  \lr{\cons}_t + \gradientX \cdot \fphys  = \Null, \hspace{0.5cm} \fphys = \Fc - \Fv,
  \label{eq:theory:nse}
\end{align}
where $\cons=\arr{\rho, \rho \vel, \rho e} \in \R^{\nvar}$, $\nvar=5$, is the vector of conserved variables, consisting of the density $\rho$, the velocity
vector $\vel=\arr{\vel[1],\vel[2],\vel[3]} \in \R^3$, and the total energy $e$ per unit mass, and $\fphys \in
\R^{3\times\nvar}$ denote the physical fluxes, comprised of the convective flux $\Fc$ and diffusive flux $\Fv$, given as
\begin{align}
  \Fc =& \arr{\rho \vel, \rho \vel \otimes \vel + \p \mathbf{I}, (\rho e + \p) \vel},\\
  \Fv =& \arr{0, \stress, \stress \cdot \vel + \conductivity \nabla T},
  \label{eq:theory:fluxes}
\end{align}
with the unit tensor $\mathbf{I} \in \R^{3 \times 3}$, the temperature $T$, the thermal conductivity $\conductivity$ and the pressure $\p$.
Following Stokes' hypothesis, which assumes that the bulk viscosity is zero, the viscous stress tensor reduces to $\stress =
\dynvisc (\lr{\nabla \vel}^\transpose + \nabla \vel - \frac{2}{3} \lr{\nabla \cdot \vel}\mathbf{I}) \in \R^{3 \times 3}$ for a Newtonian fluid, where $\dynvisc$ denotes the dynamic viscosity.
The heat flux is modeled according to Fourier's hypothesis.
The NSE are closed by the equation of state of a calorically perfect gas.
Thus, $\lambda=\frac{c_p \dynvisc}{\mathrm{Pr}}$ with the Prandtl number $\mathrm{Pr}$ and the specific heat at constant pressure $c_p$ of ambient air.
If not stated otherwise, the fluid has a temperature-dependent viscosity, $\dynvisc=\dynvisc(T)$, which follows Sutherland's law~\cite{Sutherland1893}.

\subsection{Entropy variables}
\label{sec:theory:entropy}

The NSE are equipped with an entropy/entropy flux pair~\cite{Hughes1986}, given as
\begin{align}
  \lr{v,f^s} = \lr{-\frac{\rho \fphysentropyref}{\gamma-1},-\frac{\rho \fphysentropyref \vel}{\gamma-1}}
\end{align}
with the thermodynamic entropy $\fphysentropyref=\log(p\rho^{-\gamma})$.
The entropy variables are the Jacobian of the (mathematical) entropy with respect to the conservative variables, defined as
\begin{align}
  \entropy(\cons) \coloneq \fracp{v}{\cons} =  \arr{\frac{\gamma-\fphysentropyref}{\gamma-1}-\frac{\rho}{2p} \abs{\vel}^2,\frac{\rho
  \vel}{p},-\frac{\rho}{p}}.
\end{align}
The conservative variables can be expressed by the entropy variables as
\begin{align}
  \cons(\entropytilde) = \arr{- (\rho \epsilon) \entropytilde[5], (\rho \epsilon) \entropytilde[2:4], (\rho \epsilon)
  \lr{1-\frac{\abs{\entropytilde[2:4]}^2}{2\entropytilde[5]}}}
\end{align}
with $\entropytilde=(\gamma-1)\entropy$ and the specific internal energy $\rho \epsilon$ defined in terms of the entropy variables as
\begin{align}
  \rho \epsilon = \lr{\frac{(\gamma-1)}{(-\entropytilde[5])^\gamma}}^{1/(\gamma-1)} \exp{\lr{\frac{-\fphysentropyref}{\gamma-1}}}, \
  \fphysentropyref = \gamma - \entropytilde[1] + \frac{\abs{\entropytilde[2:4]}^2}{2 \entropytilde[5]},
\end{align}
where a well defined entropy requires positivity of density and pressure.
Multiplying~\cref{eq:theory:nse} with the entropy variables and integration over $\Omega$ allows to derive the entropy
inequality for the compressible Navier--Stokes equations, see e.g.~\cite{Ray2017,Chan2018,Chan2022b} for more details.

\section{Entropy stable DGSEM on hexahedral meshes}
\label{sec:methods}
In the following, we briefly discuss the numerical treatment of the governing equations.
The open-source framework FLEXI\footnote{www.flexi-project.org}~\cite{Krais2021,Kempf2024} is used as a solver which includes the numerical methods mentioned below.

\subsection{Weak form}
Following the method of lines approach, the conservation equations are discretized in space by the discontinuous Galerkin spectral element method.
The computational domain $\Omega$ is tessellated into non-overlapping hexahedral
elements $\mathcal{K}_c$ with possibly curved faces, approximated in a tensor-product manner by one-dimensional Lagrange polynomials of degree $\ppngeo$, see~\cite{Hindenlang2012} for further details.
To obtain the DGSE formulation of~\cref{eq:theory:nse}, the following steps are necessary.
First, each element $\mathcal{K}_c \in \mathcal{T}_h$ is mapped onto the reference space $E=[-1,1]^3$ with
the mapping $\boldsymbol{\chi}: E \to \mathcal{K}_c, \refpos \mapsto \pos$.
Here $\pos=\arr{\pos[1],\pos[2],\pos[3]} \in \Omega$ and $\refpos=\arr{\refpos[1],\refpos[2],\refpos[3]} \in E$ are the coordinates in physical and computational space, respectively.
The Jacobian matrix of this mapping is $\Jm = \lr{\gradientXI \boldsymbol{\chi}} \in \R^{3\times 3}$ with $\J=\det\Jm$ as the corresponding determinant.
Then, the element-local solution $\cons$ is approximated by a polynomial representation using the tensor-product of one-dimensional nodal Lagrange basis functions of degree $\ppn$,
\begin{align}
  \cons\approx\consh(\refpos,t)=\inte(\cons) = \sum_{i,j,k=0}^\ppn \hat{\cons}_{ijk}(t)
  \ell_i(\refpos[1])\ell_j(\refpos[2])\ell_k(\refpos[3]) \in \R^{\nvar\times\nq}
\end{align}
with Gauss--Lobatto (GL) or Gauss--Legendre (G) nodes as interpolation points, the nodal degrees of freedom $\hat{\cons}(t)$, and the number of interpolation nodes in the volume $\nq \in \N$.
In the following, the subscript $h$ in $\consh$ is omitted for reasons of clarity.
Then, the resulting equations are projected in the local $L_2$ space to the surface spanned by the polynomials $\testfunc \in
\P(E,\R^{\nq})$ of order $\ppn$. Due to the Galerkin approach, $\testfunc_{ijk}=\lagrange_i(\refpos[1])\lagrange_j(\refpos[2])\lagrange_k(\refpos[3])$ with $i,j,k=0,\ldots,\ppn$.
Integration by parts leads to the weak form, written as
\begin{align}
  \projE{\J (\cons)_{t},\testfunc(\refpos)} - \projE{\fphysref,
  \nabla_{\refpos} \testfunc(\refpos)}
    +& \intpE{(\fphysref\cdot\normalvecref)^\ast \testfunc(\refpos)}  = \Null,
  \label{eq:dgsem_weak}
\end{align}
where $\projE{\cdot,\testfunc}$ denotes the volume integral over $\refraum$ and $\normalvecref \in \R^3$ is the unit normal vector in the unit reference space $\refraum$.
The contravariant fluxes are given as
\begin{align}
  \fphysref &= \M^\transpose \fphys, \\
  (\M)^\transpose &=\lr{\adj~\Jm}^\transpose \otimes \Iunit_{\nvar},
\end{align}
with the metrics terms $\M$ defined as the adjoint $\adj$ of $\Jm$, the contravariant fluxes $\fphysref \in \R^{3\times \nvar \times\nq}$ and the identity matrix $\Iunit_{\nvar} \in \R^{\nvar\times\nvar}$.
The metric terms are approximated by the conservative curl form~\cite{Kopriva2006} to guarantee that the metric identities, $\divXI
\adj~\Jm = \mathbf{0}$, hold on the discrete level since interpolation and differentiation only commute if the discretization error is zero, otherwise
$\mathbb{I}^{\ppn}(\mathbf{q}') \neq (\mathbb{I}^{\ppn}(\mathbf{q}))'$, cf.~\citep[22]{Kopriva2009}.
Neighboring elements are weakly coupled via the numerical flux $\fphys^\ast$ normal to the element faces $\{\Gamma^{\ifa}\}_{\ifa \in
\{1:\nf\}}$ with $\nf$ being the number of element surfaces, resulting in
\begin{align}
  (\fphysref \cdot\normalvecref^{(\ifa)})^\ast = (\fphys^{(\ifa)} \cdot\normalvec^{(\ifa)})^\ast \J^{(\ifa)} =
    f^\ast(\cons^{(\ifa),+} , \cons^{(\ifa),-}; \normalvec^{(\ifa)}) \J^{(\ifa)} = f^\ast_{\ifa}.
\end{align}
The outward-pointing physical normal vector $\normalvec \in \R^3$ is computed via Nanson's formula
\begin{align}
  \abs{\lr{\adj~\Jm}^{\transpose} \normalvecref^{(\ifa)}} \normalvec^{(\ifa)} = \lr{\adj~\Jm}^{\transpose} \normalvecref^{(\ifa)}
\end{align}
with the surface element $\J^{(\ifa)} = \abs{\lr{\adj~\Jm}^{\transpose} \normalvecref^{(\ifa)}}$.

The numerical flux function $f^\ast$ is approximated by Roe's numerical flux with the entropy fix by~\citet{Harten1983b}.
The approximate solutions on the left $\cons^{(\ifa),+}$ and right $\cons^{(\ifa),-}$ to an element interface $\Gamma^{\ifa}$ are given as (for a face which normal vector points in negative $\refpos[3]$-direction)
\begin{align}
  {\cons}_{ij}^{\ifa=-\refpos[3]} = \sum_{k=0}^{\ppn} {\cons}_{ijk} \lagrange_k(-1)
\end{align}
due to the tensor-product nature of the interpolation.

The collocation property is exploited for the numerical integration of~\cref{eq:dgsem_weak} by the use of $(\ppn+1)^3$
Gauss--Lobatto or Gauss--Legendre quadrature points as interpolation nodes for $\cons$, leading to the semi-discrete form given as
\begin{align}
  \Mm \J          (\cons)_t + \sum_{p=1}^{d} \Mm \Dm^\transpose \left({\M_p^{\transpose}}\right) \fphys_p(\cons(\refpos)))
                            + \sum_{p=1}^{d} \Vm_f^\transpose \Bm \left[ \begin{matrix}
    f^\ast_{\ifa=-\refpos[p]} \\
    f^\ast_{\ifa=+\refpos[p]} \\
  \end{matrix} \right] = \Null
  \label{eq:dgsem_weak_discrete}
\end{align}
with the polynomial derivative matrix $\Dm$, the mass matrix $\Mm$, the surface Vandermonde $\Vm$ and the surface weights $\Bm$
\begin{align}
  \Mm &= \diag(\w_{i})            , \ \Dm_{ij} = \ell'_j(\refpos_i), \ \Bm = \diag(-1,1), \\
  \Vm_f &=\left[ \begin{matrix} \lagrange_0(-1) & \ldots & \lagrange_{\ppn}(-1) \\
                                \lagrange_0(+1) & \ldots & \lagrange_{\ppn}(+1) \end{matrix} \right], \ i,j,k=0,...,\ppn.
\end{align}
The quadrature weights are denoted as $\w \in R^{\ppn+1}$.

The superior integration accuracy of DGSEM on Gauss–Legendre nodes, with a value of $2\ppn+1$, is a notable advantage over Gauss–Lobatto nodes, which only achieve $2\ppn-1$.
Furthermore, it is imperative to note that the utilisation of a diagonal mass matrix is exact for Gauss–Legendre nodes. Conversely, for Gauss–Lobatto nodes, this method is only approximate, which consequently leads to mass lumping.
However, a diagonal mass matrix is necessary to obtain a diagonal-norm SBP operator which is the basis for the entropy stable
DGSEM on Gauss--Lobatto nodes~\cite{Gassner2014}.
In contrast, Gauss--Lobatto nodes exhibit superior computational efficiency due to their less restrictive time
step~\cite{Kopriva2010} and lower computational demands, e.g., due to the absence of the entropy projection~\cite{Ranocha2021}.

\subsection{Viscous fluxes}
The NSE require the gradients of the primitive variables $\nabla \cons^{\text{prim}}$ to evaluate the viscous fluxes.
In this work, the BR1 (lifting) scheme~\cite{Bassi1997} is used to approximate the gradients of $\cons^{\text{prim}}$ such that
the additional set of equations
\begin{align}
  \mathbf{g} = \nabla_x \cons^{\text{prim}}
\end{align}
has to be fulfilled, where $\mathbf{g}$ are the lifted gradients.
The solution of this additional set of equations can be obtained similarly as above by deriving the weak form and applying
the DGSEM, as described in~\cite{Krais2021}.
The semi-discrete form of the final equation can be written as
\begin{align}
  \Mm \J          (\cons)_t &+ \sum\limits_{p=1}^{3} \Mm \Dm^\transpose \M_p^{\transpose}
  \left(\fphys^c_p(\cons) + \fphys^v_p(\cons, \mathbf{g}) \right) \nonumber \\
                             &+ \sum\limits_{p=1}^{3} \Vm_f^\transpose \Bm \left[ \begin{matrix}
                            f^\ast_{\ifa=-\refpos[p]}+\J^{(\ifa=-\refpos[p])}\avg{{\fphys^v}^{(\ifa=-\refpos[p])}} \\
                            f^\ast_{\ifa=+\refpos[p]}+\J^{(\ifa=+\refpos[p])}\avg{{\fphys^v}^{(\ifa=+\refpos[p])}} \\
  \end{matrix} \right] = \Null
\end{align}
with the averaging operator $\avg{\cdot}_{ij} = [(\cdot)_i+(\cdot)_j]/2$.

The solution is advanced in time by an explicit low-storage Runge--Kutta (RK) fourth-order accurate scheme with five stages
(RK4-5), proposed by~\citet{Carpenter1994}. If not stated otherwise, in this work a $\mathrm{CFL}$ (Courant, Friedrichs and Lewy) number of $\mathrm{CFL}=0.9$ is set, the same applies to the viscous time step restriction.
It is important to note that Gauss--Lobatto nodes permit a larger time step than Gauss--Legendre nodes for a given polynomial degree and Runge--Kutta scheme~\cite{Chan2016}.

\subsection{\label{sec:methods:ESDGSEM_GL}Entropy stable DGSEM on Gauss--Lobatto nodes}

In this paper, the split formulation of the DGSEM is utilized to alleviate numerical stability issues and
mitigate aliasing errors due to the approximation of the nonlinear (convective) flux by $\inte(\cdot)$, see e.g.~\cite{Gassner2016,Flad2017} for more details.
The idea is based on the work of Fisher and Carpenter~\cite{Fisher2013,Carpenter2014}, which have
proven that the summation-by-parts (SBP) property and a two-point flux which satisfies Tadmor's discrete entropy
condition~\cite{Tadmor1987} enables the derivation of entropy conservative high-order schemes on finite domains.
Based on this work,~\citet{Gassner2016} demonstrated that the SBP property of Gauss--Lobatto
nodes and a suitable two-point flux function allows to construct an entropy or kinetic energy stable
DGSEM on Gauss--Lobatto nodes. %

Finally, with the SBP property, $\Qm+\Qm^\transpose=\Vm_f^T\Bm$ with $\Qm=\Mm\Dm$, and an adequate two-point flux, the entropy stable DGSEM on Gauss--Lobatto nodes is written as
\begin{align}
  \Mm \J          (\cons)_t
  &+ \sum_{p=1}^{d} \lr{2 \mathbf{Q} \circ \fcontsharp_p} \mathbf{1} + \Mm \Dm^{\transpose} \fphys^v_p \\
  &+ \sum_{p=1}^{d} \Vm_f^\transpose \Bm \left[ \begin{matrix}
    f^\ast_{\ifa=-\refpos[p]} - \fcontsharp(\cons^{(\ifa=-\refpos[p])})+\J^{(\ifa=-\refpos[p])}\avg{{\fphys^v}^{(\ifa=-\refpos[p])}} \\
    f^\ast_{\ifa=+\refpos[p]} - \fcontsharp(\cons^{(\ifa=+\refpos[p])})+\J^{(\ifa=+\refpos[p])}\avg{{\fphys^v}^{(\ifa=+\refpos[p])}} \\
  \end{matrix} \right] = \Null \nonumber
  \label{eq:ESDGSEM_GL}
\end{align}
with the vector of all ones $\mathbf{1}$, the contravariant two-point fluxes $\fcontsharp$ in the volume and $\fcontsharp(\cons^{(\ifa)})$ on the surface, defined as
\begin{align}
  {\fcontsharp}_{ij} &= \avg{\M^\transpose}_{ij} f^{\#} ({\cons}_i, {\cons}_j), \ i,j=0,...,\nq, \\
  {\fcontsharp}^{(\ifa)}_{mn} &= \avg{\M^\transpose}_{mn} f^{\#} ({\cons}_m, {\cons}_n), \ m=0,...,\nq, \ n=1,...,(\ppn+1)^{d-1}.
\end{align}
Depending on the specific form of $f^{\#}$, desired criteria can be enforced on the discrete level, kinetic energy
preservation (with the definition of~\citet{Jameson2008}) or entropy
conservation (in the sense of~\citet{Tadmor1987}).

\subsection{\label{sec:methods:ESDGSEM_G}Entropy stable DGSEM on Gauss--Legendre nodes}
Recently, the entropy stable DG scheme was extended to Gauss--Legendre nodes by exploiting a generalized SBP
property~\cite{Chan2018,Chan2019} and it was also proven that it can be written in a flux differencing form~\cite{Mateo-Gabin2023}.
Following~\citet{Chan2018}, the matrix formulation for the entropy-stable DGSEM on Gauss--Legendre nodes can be derived with the generalized SBP operator
\begin{align}
  \Qm =
  \left(\begin{matrix}
    2\Mm \Dm - \Vm_f^\transpose \Bm \Vm_f & & \Vm_f^\transpose \Bm \\
    -\Vm_f \Bm                            & & \Bm
  \end{matrix}\right)
  \label{eq:SBP_G}
\end{align}
and the entropy variables $\entropy$, resulting in
\begin{align}
  \Mm \J          (\cons)_t
  &+ \sum_{p=1}^{d} \left[\begin{matrix} \mathbf{I} \ \Vm_f^{\transpose} \end{matrix} \right] \lr{\mathbf{Q} \circ
  \fcontsharp_p} \mathbf{1}
  + \Mm \Dm^{\transpose} \fphys^v_p \\
  &+ \sum_{p=1}^{d} \Vm_f^\transpose \Bm \left[ \begin{matrix}
    f^\ast_{\ifa=-\refpos[p]} - \fcontsharp(\qentropyproj^{(\ifa=-\refpos[p])})+\J^{(\ifa=-\refpos[p])}\avg{{\fphys^v}^{(\ifa=-\refpos[p])}} \\
    f^\ast_{\ifa=+\refpos[p]} - \fcontsharp(\qentropyproj^{(\ifa=+\refpos[p])})+\J^{(\ifa=+\refpos[p])}\avg{{\fphys^v}^{(\ifa=+\refpos[p])}} \\
  \end{matrix} \right] = \Null.\nonumber
  \label{eq:ESDGSEM_G}
\end{align}
Here the contravariant fluxes are defined as above and the entropy projected variables on the surface are
\begin{align}
  \qentropyproj = \cons(\entropyvarsurf), \hspace{1cm}
  {\entropyvarsurf}_{ij}^{\ifa=-\refpos[3]} = \sum_{k=0}^{\ppn} {\entropy(\cons)}_{ijk} \lagrange_k(-1).
\end{align}

\subsection{\label{sec:methods:shock_caputring}Shock capturing}

In this work, the shock capturing is based on a convex blending of a low-order FV scheme, see
e.g.~\cite{Sonntag2017}, here a second-order total variation diminishing FV subcell approach, with the higher-order DGSE operator~\cite{Hennemann2021}.
The hybrid operator can be written as
\begin{align}
  (\J \cons)_t = \FValpha \lr{(\J \cons)_{t}}^{\mathrm{FV}} + (1-\FValpha) \lr{(\J \cons)_{t}}^{\mathrm{DG}}.
  \label{eq:semi_discrete_DGFV}
\end{align}
The authors in~\cite{Hennemann2021} proposed a modified version of the modal shock indicator of~\cite{Persson2006} and introduced two
tuning parameters, to predict the blending function $\FValpha \in [0,1]$.
The product of the collocated density and pressure are utilized as an indicator variable.
The FV subcell operator is discretized similarly to the DG operator on Gauss--Lobatto or Gauss--Legendre nodes and
interpreted the nodal solution at these points as integral mean values of the FV method.
The advantage of this approach is that only the volume term has to be blended, as long as the solution at the element interfaces is
equivalent for both schemes.
This is trivially fulfilled for a first-order FV scheme defined on Gauss--Lobatto nodes, but has to be enforced for a second-order FV scheme and/or Gauss--Legendre nodes, either through an additional blending of the surface term, cf.~\cite{Rueda-Ramirez2022}, or by using a zero slope in the outer subcells (when on Gauss--Lobatto nodes).
To ensure a unique $\FValpha$ on each element interface shared by two adjacent DG elements, $\FValpha_f$ is computed as the maximum between the element to the left, $\FValpha^-$, and the right, $\FValpha^+$, of the corresponding DG interface, leading to $\FValpha_{f}=\max(\FValpha^-,\FValpha^+)$, following~\cite{Rueda-Ramirez2022}. A piecewise linear reconstruction is applied to obtain the second-order total variation diminishing FV subcell scheme using the
gminmod limiter~\cite{Leer1977}.

\section{Numerical results}
\label{sec:application}

In the following section, a detailed comparison of the entropy-stable DGSEM (ESDGSEM) on Gauss–Legendre (DGSEM-G) and Gauss–Lobatto
(DGSEM-GL) nodes will be
presented, starting with a generic manufactured solution and progressing to increasingly complex examples of the Taylor-Green vortex and
the wall-bounded turbulent flow in a 3D diffuser.
The manufactured solution allows to illustrate the convergence properties of the ESDGSEM.
The Taylor-Green vortex is a popular test case for turbulent flows to compare the accuracy, stability, and efficiency of
the ESDGSEM-G and ESDGSEM-GL. Two versions of the Taylor-Green vortex are considered: A weakly incompressible case to assess the
approximation quality of small scale turbulence. The compressible case allows to evaluate the robustness and accuracy for
shock-turbulence interactions.
Finally, this section concludes with a comparison of the approximation quality for wall-bounded turbulent flow using a 3D diffuser
as a test case. Two configurations of this diffuser are considered: First, a weakly compressible case, for which reference data are
available, and a compressible case. This allows to compare the convergence properties, accuracy, and efficiency of the ESDGSEM-GL
and ESDGSEM-G for a more complex application.

If not stated otherwise, the entropy stable flux of~\citet{Chandrashekar2013} is utilized.
The shock capturing procedure is only applied for the compressible Taylor-Green vortex.

\subsection{Experimental order of convergence}
\label{sec:EOC}
\begin{figure}
  \includegraphics[width=\linewidth]{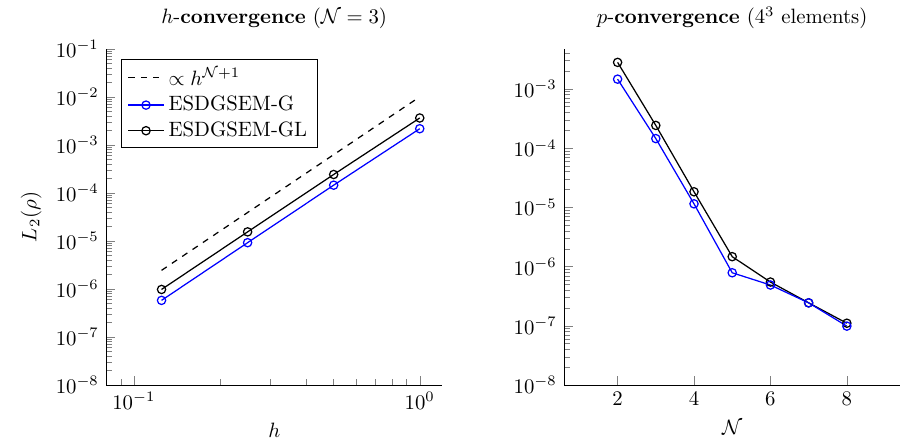}
  \caption{Validation of the spatial discretization. Left: $h$-convergence for $\ppn=3$. Grid sequence ranges from $2^3$ up to
  $32^3$. Right: $p$-convergence on a $4^3$ grid with $\ppn \in [2,8]$.}
  \label{fig:application:conv}
\end{figure}

First, briefly the $p$- and $h$-convergence of the ESDGSEM on Gauss--Legendre and Gauss--Lobatto nodes is discussed using the
method of manufactured solutions, cf.~\cite{Gassner2009} for further details.
Following~\cite{Hindenlang2012}, the exact solution is assumed to be of the form
\begin{align}
  \rho = 2+0.1\sin(2\pi(\pos[1]+\pos[2]+\pos[3]-t)), \ \rho \vel = \rho, \ \rho e = (\rho)^2.
\end{align}
For the $p$-convergence study, the computational domain $\Omega \in [0,1]$ is discretized by $4^3$ elements and the polynomial
degree was varied, while for the $h$-convergence study, $\Omega$ was discretized using $4^3$ to $32^3$ elements and $\ppn=3$.
In addition, the domain was sinusoidally deformed according to~\cite{Minoli2011} to assess the convergence properties for curved DGSEM.
The chosen error norm is the discrete $L_2$ error of the density.
The results in~\cref{fig:application:conv} highlight the spatial convergence properties of the ESDGSEM-G and ESDGSEM-GL, which
both achieve the expected order of convergence.
As expected, the ESDGSEM-GL is less accurate due to the lower integration accuracy of $2\ppn -1$ compared to Gauss--Legendre with an integration accuracy of $2\ppn+1$.

\subsection{Taylor-Green vortex}
\label{sec:tgv}

The Taylor-Green vortex (TGV)~\cite{Taylor1937} is a popular validation case for turbulence.
The corresponding analytical initial conditions are given as
\begin{align}
  \vel(\pos,t=0) &= \left[ \begin{matrix} u_0\sin(\nicefrac{\pos[1]}{L})\cos(\nicefrac{\pos[2]}{L})\cos(\nicefrac{\pos[3]}{L}) \\
                                         -u_0\cos(\nicefrac{\pos[1]}{L})\cos(\nicefrac{\pos[2]}{L})\cos(\nicefrac{\pos[3]}{L}) \\ 0
                                     \end{matrix} \right], \\
  p(\pos,t=0) &= p_0 + \nicefrac{\rho_0
      u_0^2}{16}\lr{\cos(\nicefrac{2\pos[1]}{L})+\cos(\nicefrac{2\pos[2]}{L})}\lr{2+\cos(\nicefrac{2\pos[3]}{L})}, \nonumber
  \label{eq:TGV_init}
\end{align}
on a computational domain of size $\Omega = [0,2\pi L]^3$ with periodic boundary conditions.
Here, $L=1$ denotes the characteristic length, $u_0$ is the magnitude of the initial velocity fluctuations and $\rho_0$ is the
reference density.
The initial pressure $p_0$ is chosen consistent to the prescribed Mach number $\mathrm{M}_0 = u_0 \sqrt{\nicefrac{\rho_0}{\gamma
p_0}}$. The desired Reynolds number, $\mathrm{Re} = \nicefrac{\rho_0 u_0 L}{\mu_0}$ is obtained by adjusting the dynamic viscosity $\mu_0$.
The density is obtained from the equation of state of an ideal gas $\rho = \nicefrac{p}{RT}$ with $T(\pos,t=0)=T_0$.

To assess the accuracy of numerical schemes for the TGV, three common metrics are the instantaneous kinetic energy $e_k$
together with the solenoidal $\epsilon_S$ and dilatational $\epsilon_D$ component of the viscous dissipation rate.
\begin{align}
  e_k        &= \frac{1}{2 \rho_0 u_0^2 \abs{\Omega}} \int_{\Omega} \rho \vel \cdot \vel~d\Omega, \\
  \epsilon_S &= \frac{L^2}{\mathrm{Re} u_0^2 \abs{\Omega}} \int_{\Omega} \frac{\mu(T)}{\mu_0} \boldsymbol{\omega} \cdot \boldsymbol{\omega}~d \Omega, \\
  \epsilon_D &= \frac{4L^2}{3 \mathrm{Re} u_0^2 \abs{\Omega}} \int_{\Omega} \frac{\mu(T)}{\mu_0} (\nabla \cdot \vel)^2~d \Omega,
\end{align}
respectively. Here, $\boldsymbol{\omega}=\nabla \times \vel$ denotes the vorticity. The solenoidal dissipation rate can be related to the amount of dissipated kinetic energy through small scale turbulence
and the dilatational dissipation rate to kinetic energy dissipation due to compressibility effects.

In the following, two versions of the TGV are considered: First, the weakly compressible TGV with $\mathrm{M}_0=0.1$, now
referred to as incompressible TGV, to evaluate the
accuracy of capturing the small scale turbulence. Second, the compressible TGV with $\mathrm{M}_0=1.25$ to additional assess the stability of the schemes.
The Reynolds number in both cases is chosen as $\mathrm{Re}=1600$.

\subsubsection{Incompressible TGV}
\label{sec:tgv_incomp}

\begin{figure}
  \centering
  \includegraphics[width=0.8\linewidth]{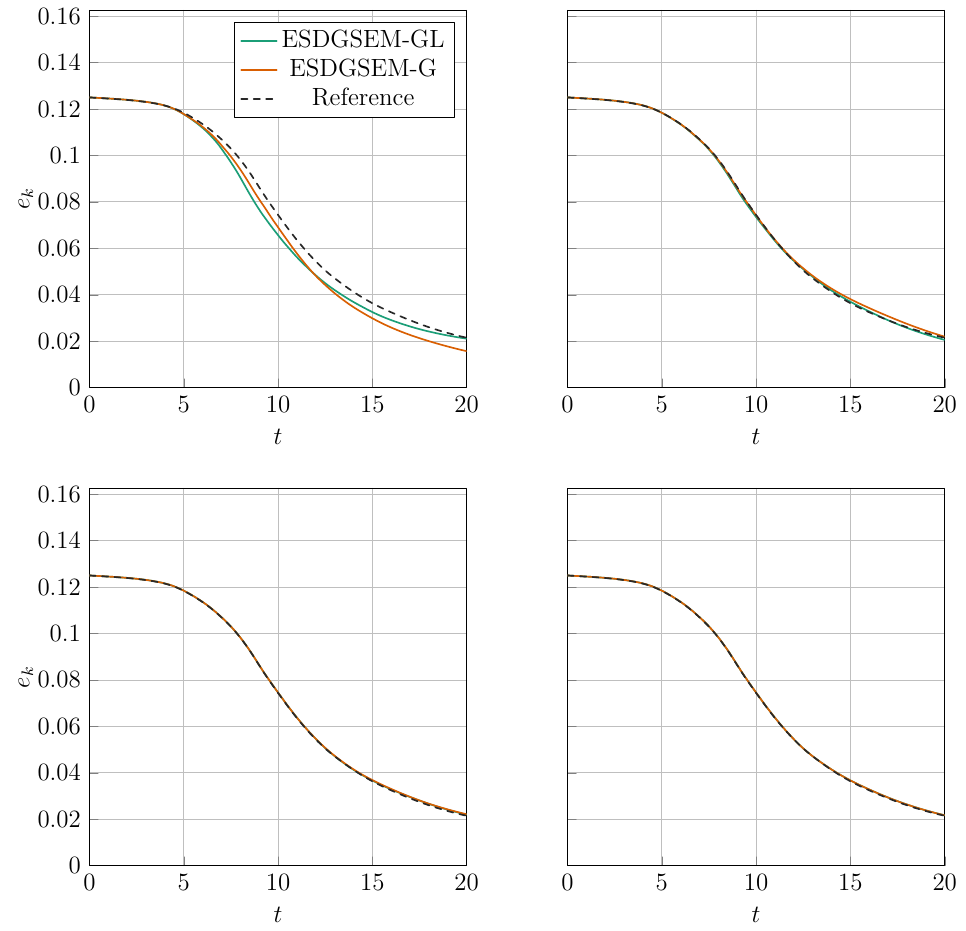}
  \caption{Temporal evolution of the instantaneous kinetic energy for $64^3$ DOF, $128^3$ DOF, $256^3$ DOF,
  and $512^3$ DOF (from left to right and top to bottom) for the incompressible TGV at $\mathrm{M}_0=0.1$. The results of~\citet{Debonis2013} serve as a reference.}
  \label{fig:tgv_incomp_ek}
\end{figure}

\begin{figure}
  \centering
  \includegraphics[width=0.8\linewidth]{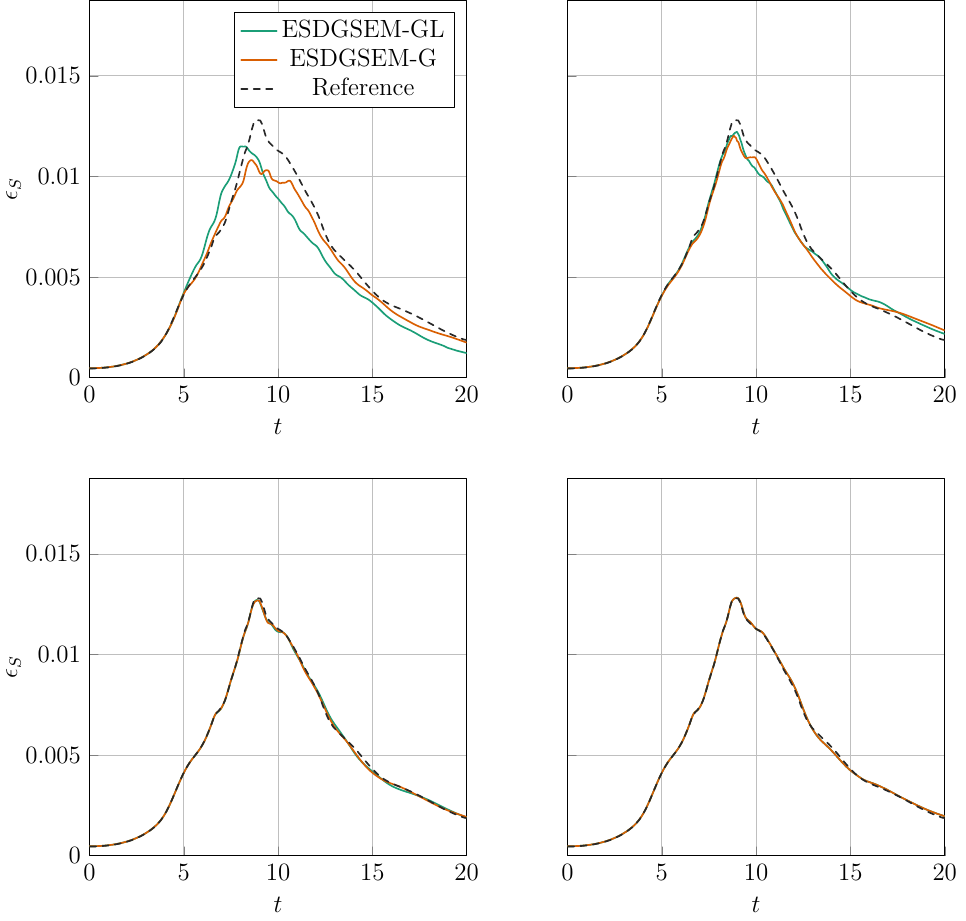}
  \caption{Temporal evolution of the solenoidal dissipation rate for $64^3$ DOF, $128^3$ DOF, $256^3$ DOF,
  and $512^3$ DOF (from left to right and top to bottom) for the incompressible TGV at $\mathrm{M}_0=0.1$. The results of~\citet{Debonis2013} serve as a reference.}
  \label{fig:tgv_incomp_es}
\end{figure}

Four different grid resolutions are considered, resulting in $64^3$ up to $512^3$ degrees of freedom (DOF) with a polynomial degree
of $\ppn=7$ and Lax-Friedrich's numerical flux function.
The results are also validated against the reference data by~\citet{Debonis2013} using the temporal evolution of the instantaneous kinetic energy and the solenoidal dissipation rate as metrics.
The convergence of the instantaneous kinetic energy and the solenoidal dissipation rate during mesh refinement to the reference is
illustrated in~\cref{fig:tgv_incomp_es} and~\cref{fig:tgv_incomp_ek} for the different configurations.
The results for the solenoidal dissipation rate show that the lower integration accuracy of ESDGSEM-GL compared to ESDGSEM-G leads to a slower convergence to the reference.
A similar behavior is visible for the kinetic energy.
Particular differences are seen in the solenoidal dissipation rate for the $64^3$ case, where the agreement of the ESDGSEM-G with the reference is far better than for the ESDGSEM-GL.
As such, ESDGSEM-G can be considered being more accurate in capturing the small scale turbulence.
It has to be noted that ESDGSEM-G takes about double the amount of computational time since more computations are necessary and the time step is more severe compared to ESDGSEM-GL, as discussed in~\cite{Kopriva2010,Ranocha2021}.

\subsubsection{Compressible TGV}
\label{sec:tgv_comp}

Recently, the TGV at $\mathrm{Re}=1600$ was extended to the compressible regime~\cite{Lusher2021,Chapelier2024}. A Mach number of
$\mathrm{M}=1.25$ is a common choice due to the emergence of shock patterns which interact with the turbulent flow. This allows the
assessment of the accuracy and stability of the ESDGSEM for shock-turbulence interactions. To address the
temperature-dependency of the viscosity in the compressible case, Sutherland's law is applied.
The shock capturing scheme introduced in~\cref{sec:methods:shock_caputring} is utilized to stabilize the flow near shocks.
It has to be noted that the investigations of the influence of different shock capturing strategies is out of the scope of this
paper.

Following~\citet{Chapelier2024}, four different grid resolutions are considering, resulting in $64^3$ up to $512^3$ DOF with a polynomial degree of $\ppn=3$ and Roe's numerical flux function with the entropy fix by~\citet{Harten1983b}.
The minimum and maximum blending coefficient, given as $(0.01,0.04)$, are chosen identically for all considered resolutions.
The results presented in~\citet{Chapelier2024} serve as a reference.
Contrary to the weakly compressible TGV, now also the dilatational dissipation rate is evaluated which gives an insight to the shock
capturing capabilities of the numerical scheme by assessing the sharpness of shocks.
In addition, the presence of oscillations in the dilatational dissipation rate can be identified as a diagnostic indicator for the detection of the Gibbs phenomenon.
This in turn can be used to assess the capability of the shock capturing procedure.
A high numerical dissipation results in a lower dilatational dissipation rate and a smoother but less accurate shock capturing. A higher value of the dilatational dissipation rate indicates a low numerical dissipation and in turn a sharper shock capturing but at the expense of spurious oscillations which could comprise numerical stability.

\begin{figure}
  \centering
  \includegraphics[width=0.95\linewidth]{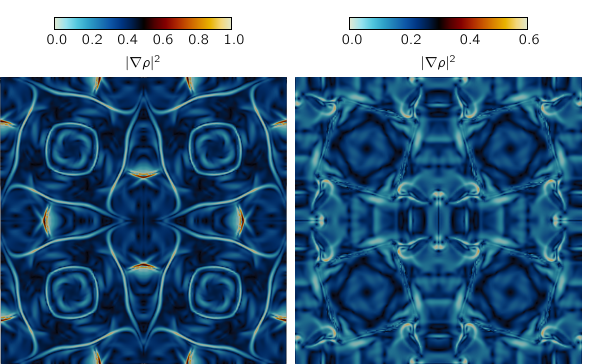}
  \caption{Instantaneous shot of the density gradient in the $\pos[1],\pos[3]$-plane at $\pos[2]=\pi$ for $t=15$ (left) and
  $t=20$ (right).}
  \label{fig:tgv_comp_schlieren}
\end{figure}

\begin{figure}
  \centering
  \includegraphics[width=0.8\linewidth]{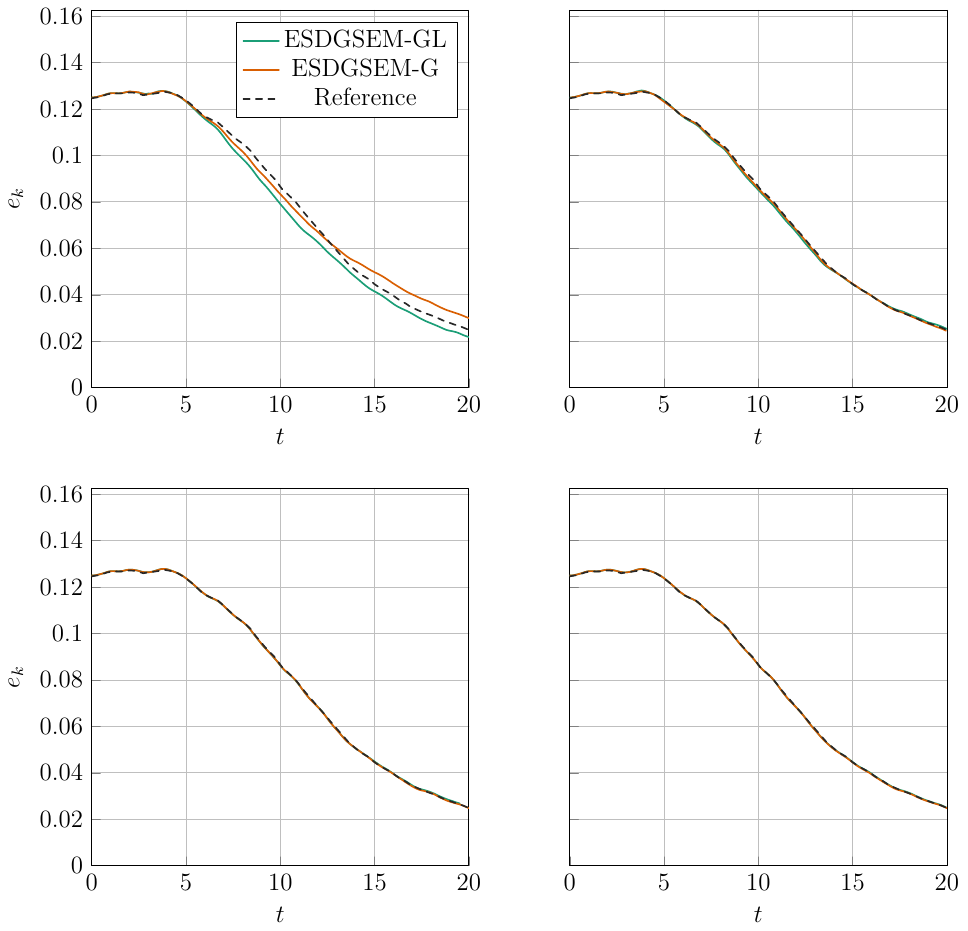}
  \caption{Temporal evolution of the instantaneous kinetic energy for $64^3$ DOF, $128^3$ DOF, $256^3$ DOF,
  and $512^3$ DOF (from left to right and top to bottom) for the compressible TGV at $\mathrm{M}_0=1.25$. The results of~\citet{Chapelier2024} serve as a reference.}
  \label{fig:tgv_comp_ek}
\end{figure}

\begin{figure}
  \centering
  \includegraphics[width=0.8\linewidth]{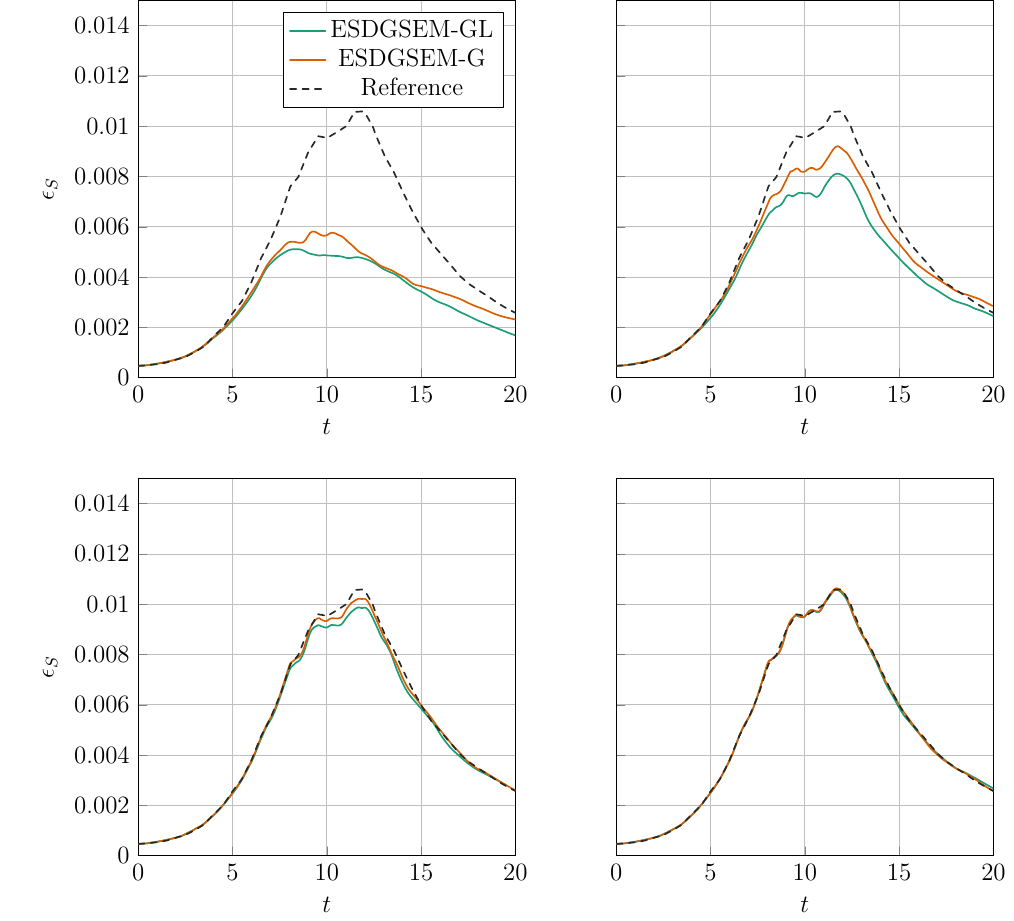}
  \caption{Temporal evolution of the solenoidal dissipation rate for $64^3$ DOF, $128^3$ DOF, $256^3$ DOF,
  and $512^3$ DOF (from left to right and top to bottom) for the compressible TGV at $\mathrm{M}_0=1.25$. The results by~\citet{Chapelier2024} serve as a reference.}
  \label{fig:tgv_comp_es}
\end{figure}

\begin{figure}
  \centering
  \includegraphics[width=0.8\linewidth]{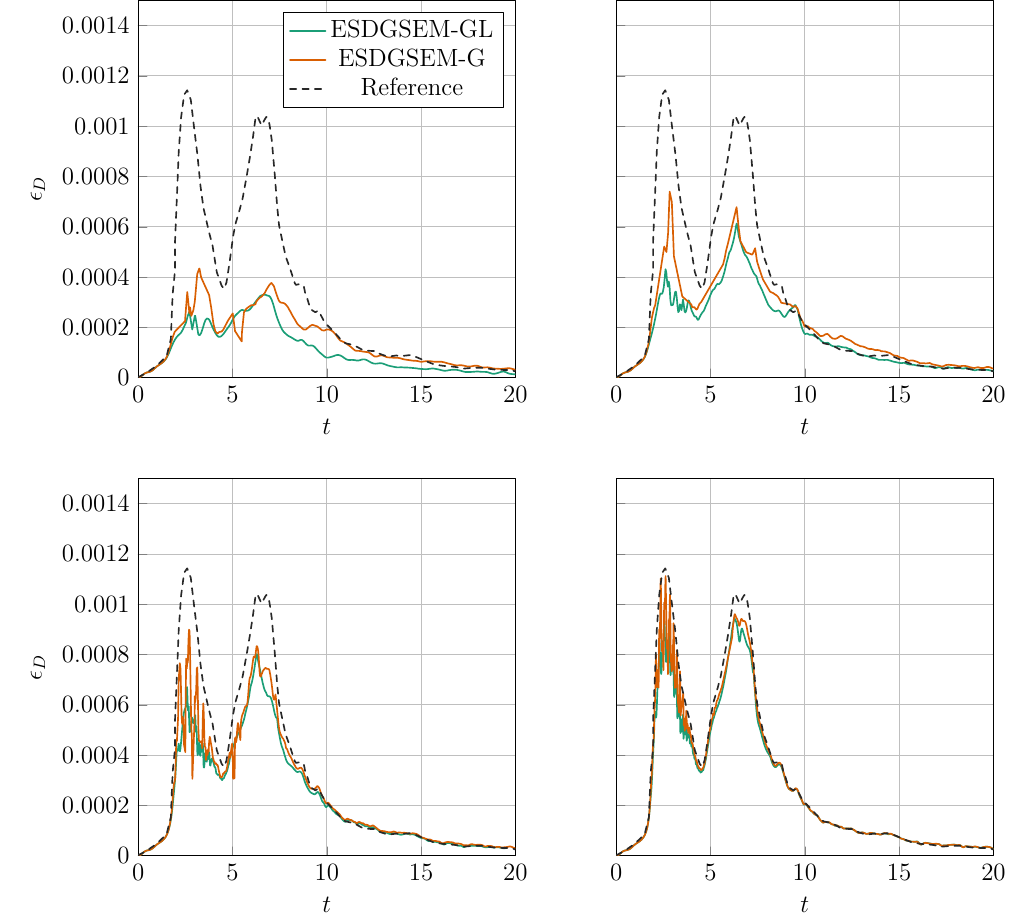}
  \caption{Temporal evolution of the dilatational dissipation rate for $64^3$ DOF, $128^3$, $256^3$ DOF,
  and $512^3$ DOF (from left to right and top to bottom) for the compressible TGV at $\mathrm{M}_0=1.25$. The results by~\citet{Chapelier2024} serve as a reference.}
  \label{fig:tgv_comp_ed}
\end{figure}

\begin{figure}
  \centering
  \includegraphics[width=0.8\linewidth]{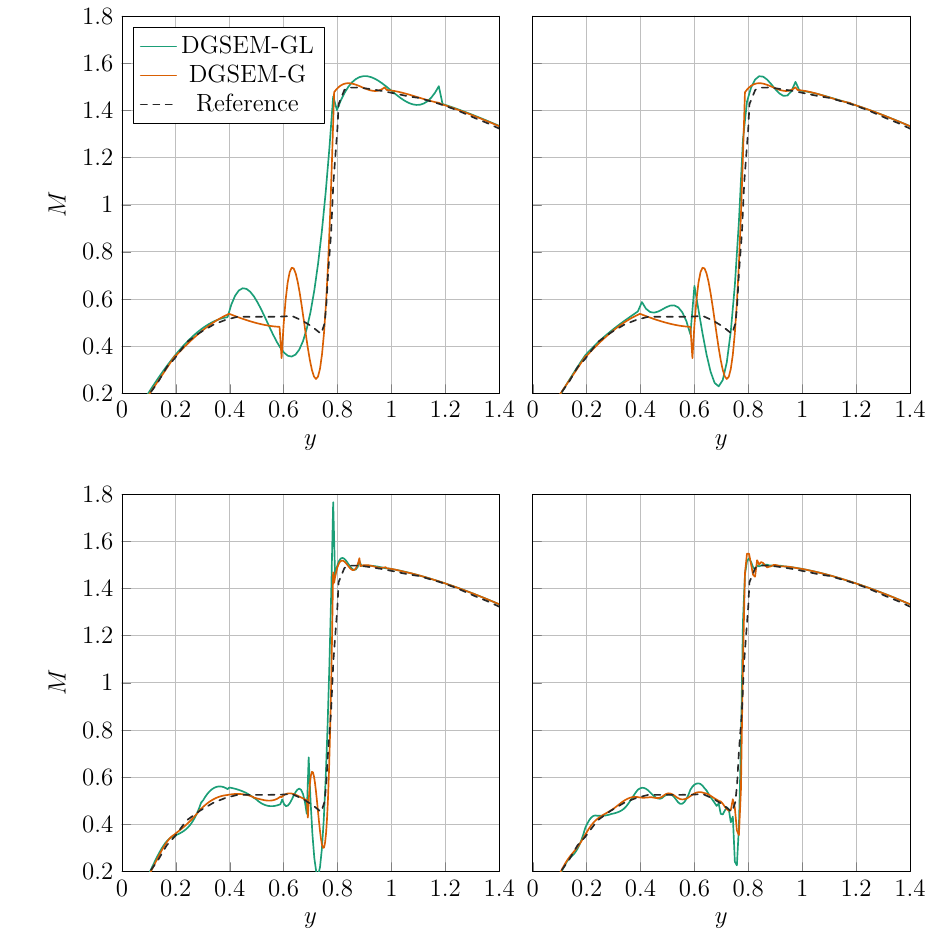}
  \caption{Mach number profiles along the $\pos[2]$ axis at $\pos[1]=\pos[3]=\pi$ and $t=2.5$ for $64^3$ DOF, $128^3$ DOF, $256^3$ DOF,
  and $512^3$ DOF (from left to right) for the compressible TGV at $\mathrm{M}_0=1.25$. The results by~\citet{Chapelier2024} serve as a reference.}
  \label{fig:tgv_comp_shock}
\end{figure}

The instantaneous flow field of the density gradient of the TGV case at $t=15$ and $t=20$ are illustrated
in~\cref{fig:tgv_comp_schlieren}, highlighting the evolution of shocks and shocklets.
The temporal evolution of the kinetic energy and solenoidal and dilatational dissipation rate
in~\cref{fig:tgv_comp_ek},~\cref{fig:tgv_comp_es}, and~\cref{fig:tgv_comp_ed}, respectively, illustrate the convergence of both schemes to the reference during grid refinement.
The results highlight that compared to the ESDGSEM-GL, the ESDGSEM-G captures small scale turbulent structures better, is overall less dissipative, and converges faster to the reference.
The Gibbs phenomenon can be identified as the underlying cause of the oscillations in the dilatational dissipation rate, which are
particularly apparent in the $256^3$ and $512^3$ cases, cf.~\cite{Chapelier2024}.
This phenomenon can be attributed to the employed blending coefficient which introduces the minimum necessary dissipation to
stabilize the numerical scheme at the expense of oscillations in the solutions.
However, it has to be noted, that the focus of this paper is not on the shock capturing procedure.
To further evaluate the shock capturing capabilities, i.e., quantify how dissipative or oscillatory either scheme is, the Mach
number profiles along the $\pos[1]=\pos[3]=\pi$ lines at $t=2.5$ are extracted for all resolutions.
The results in~\cref{fig:tgv_comp_shock} highlight that the shock is sharper for Gauss--Legendre nodes and also less oscillatory during mesh refinement compared to Gauss--Lobatto nodes.

\subsection{LES of a 3D diffuser}
\newcommand{\Ma}{M}
\newcommand{\MaDIn}{M_{D,in}}
\newcommand{\MaIn}{M_{in}}

In this section a more complex test case is investigated, a three-dimensional turbulent diffuser flow resembling a
challenging 3D corner flow problem at adverse pressure gradient.
For this, the UFR 4-16 test problem from the ERCOFTAC database \cite{cherry2008geometric,ohlsson2010direct,miro2024self}
is utilized, also known as the so called "3D Stanford Diffuser" in version 1.
This test case is of paramount importance for technical applications, given the extensive use of diffusers in numerous technical products.
In this context, it is essential to be able to predict relevant parameters, such as pressure recovery or total pressure loss, with sufficient accuracy and acceptable computing time. %

\subsubsection{Simulation setup}

\begin{figure}[t]
  \centering
  \includestandalone[width=0.9\textwidth]{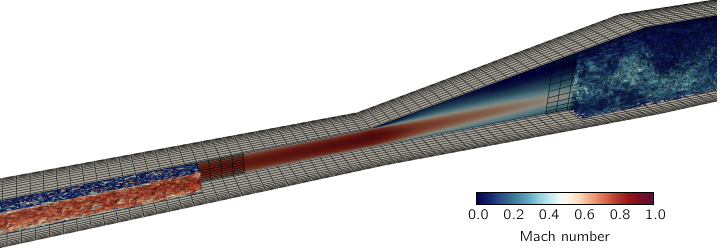}
  \caption{Illustration of the simulation domain around the diffuser section with the high-order element grid. At the center
  plane, the time-averaged Mach number distribution for the $\MaDIn=0.7$ case is depicted. Up- and
downstream of the diffuser section, the vortical structures of an instantaneous solution are indicated by iso surfaces of a constant Q-criterion value, colored by the Mach number.}
  \label{fig:diffuser_mesh}
\end{figure}

The Reynolds number studied, based on the bulk velocity $U_b$ and inflow-duct height $h$, is $\mathrm{Re}= U_b \rho h / \mu = \num{10000}$.
Despite its simple geometry, this test case results in a complex internal corner flow with 3D separation, which is difficult to
predict accurately using numerical simulation.
Following~\citet{ohlsson2010direct} and~\citet{miro2024self}, the diffuser geometry consists of a rectangular channel duct section of height
$h=1$ and width $b=3.33h$, which expands within the subsequent diffuser section of length $L=15h$ to a cross section of size $4h \times 4h $, cf.~\cref{fig:diffuser_mesh}.
This results in a diffuser angle of $11.3^\circ$ at the upper wall and $2.56^\circ$ at the side walls.
In this work, the computational domain is discretized by a structured grid. Consequently, the grid resolution along the diffuser decreases as the cross-section increases, as the number of elements in each cross-section is constant. %
The simulation domain consists of a rectangular duct upstream of the diffuser with a length of $70h$ and a triangular trip of height $0.05h$ at the upper and lower walls $2h$ after the inflow.
This duct section ensures a fully-developed turbulent flow entering the diffuser section.
Downstream of the diffuser follows a square straight diffuser-outlet section of length $12.5h$. This part is followed by a
contraction of $10h$ and a long straight duct of $5h$ to reduce the interaction of the diffuser with the outlet.

In this paper, various large-eddy simulations (LES) of this geometry are investigated: Two different Mach numbers each calculated
with the entropy stable DGSEM using Gauss–Legendre nodes (LES-G) or Gauss–Lobatto nodes (LES-GL) with $\ppn=\{5,7,11\}$.
It has to be noted that the better-resolved LES results at $\ppn=\{7,11\}$ only serve the purpose to demonstrate convergence of
the results.
The first bulk diffuser inflow Mach number considered is $\MaDIn =0.2$, resembling a weakly compressible flow, for
which a variety of reference data is available in the literature,
e.g.,~\cite{cherry2008geometric,ohlsson2010direct,miro2024self}.
The second investigated bulk diffuser inflow Mach number is $ \MaDIn = 0.7$, representing a novel extension of this test case to the compressible flow regime.
For the latter case, the absence of literature data precludes an exact validation of the compressible case, such that a
finer-resolved LES at $\ppn=7$ on Gauss--Legendre nodes serves as a reference.
However, it allows to gain insight into the performance of the ESDGSEM-GL and ESDGSEM-G schemes for complex, compressible
flows and to further verify the observations for the compressible TGV, cf.~\cref{sec:tgv_comp}.
In practical applications, coarse LES resolutions are of paramount importance due to limited computational resources and
even more complex settings.
Consequently, the ensuing discussion will focus on an in-depth comparison of the outcomes of the coarse LES-G and LES-GL simulations at $\ppn=5$.
To summarize, four different simulations will be detailed in the following: Two different Mach numbers each calculated
with LES-G or LES-GL using a polynomial degree of $\ppn=5$.

For the current investigations, the mesh consists of \num{65450} high-order elements which results in roughly \num{14} million degrees of freedom for a polynomial degree of $\ppn =5$. The resulting mesh resolution is $\pos[1]^+=46.8$, $\pos[2]^+=4.7$ and
$\pos[3]^+=14.7$.
Following~\cite{ohlsson2010direct,miro2024self}, at the channel-duct inflow, a bulk velocity of $U_b = 1$ is specified, and the remaining fluid parameters are set to match the reference
Reynolds number $\mathrm{Re} = U_b \rho h / \mu$ and a bulk Mach number of $\MaIn = 0.2$ or $\MaIn =0.45$, resulting in a diffuser
inflow bulk Mach number of $ \MaDIn =0.208$ and $ \MaDIn =0.703$, respectively.
This increase of the bulk Mach number along the rectangular duct section, see~\cref{fig:diffuser_mesh}, is caused by the total pressure drop.
At the outflow, a time-dependent pressure is chosen to ensure that the total pressure loss along the diffuser is consistent
to the experimental data~\cite{cherry2008geometric}.
All remaining boundary conditions are treated as adiabatic walls.
Further details on the setup are given in, e.g.,~\cite{ohlsson2010direct,miro2024self}.
All four simulations are initialized using a precursor simulation at $\ppn = 2$ for $ \num{1000}~T^*$ characteristic time steps, where $T^* = tU_b/h$.
At $\ppn = 5$, the simulation is further extended for $\num{1000}~T^*$ to allow the flow to converge to a statistical equilibrium state. Then, turbulent statistics are collected for $400 ~T^*$, corresponding to 26 diffuser flow-through times $T_D^* = tU_b/L$.

Following~\cite{ohlsson2010direct,miro2024self}, common metrics for the 3D diffuser flow are mean values, including the mean center
plane velocity profile, pressure coefficient, and pressure recovery. These metrics are utilized to evaluate the accuracy and convergence properties.
In addition, higher statistical moments such as Reynolds stresses are frequently used as metrics to assess the resolution capabilities of small-scale turbulence.

\subsubsection{Weakly compressible diffuser flow}
\label{sec:weaklycomprdiff}

\begin{figure}[tbp]%
  \centering
  \includestandalone[width=\linewidth]{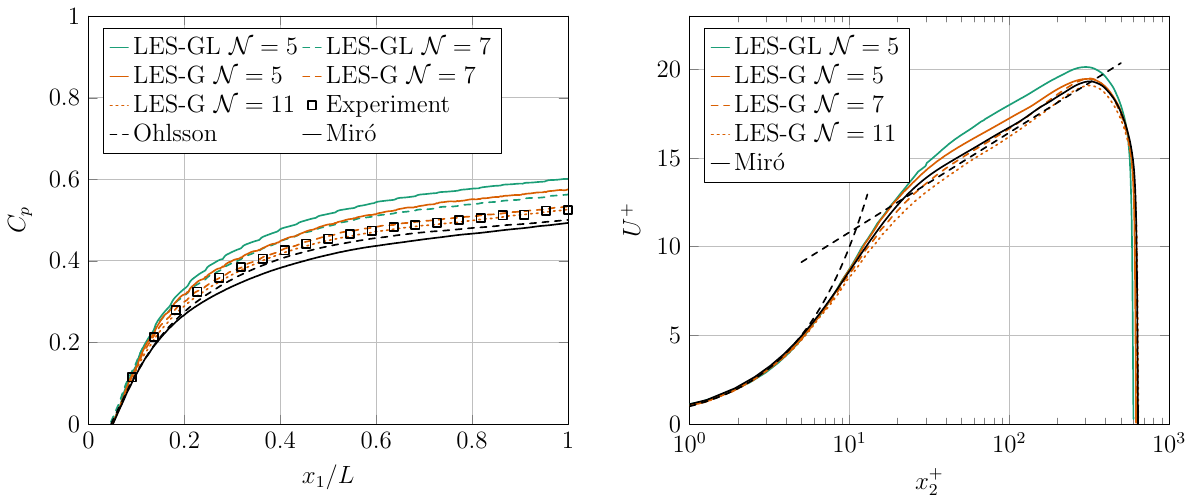}
  \caption{Comparison of the LES using Gauss--Legendre nodes (LES-G) to Gauss--Lobatto nodes (LES-GL) with $\ppn=5$ for a diffuser
    inflow bulk Mach number of $\MaDIn=0.2$.
    Mesh convergence is demonstrated using the LES-G simulation at $\ppn =7$ and $\ppn =11$.
    Left: Pressure coefficient on the bottom wall $\pos[2]/h=0$ at $\pos[3]/b=0.5$.
    The experimental data of~\citet{cherry2008geometric} and the numerical results of~\citet{ohlsson2010direct} (Nek5000)
  and~\citet{miro2024self} (Alya) serve as a reference.
    Right: Mean center plane velocity profile at $2h$ upstream of the diffuser inflow.
    The log law $\log(\pos[2]^+)/\kappa+B$ (dashed straight line) is calculated using $\kappa = 0.41$ and $B=5.2$.
  }
  \label{fig:diffuser_incomp_cp_umean}
\end{figure}

First, the convergence of the LES for the weakly compressible diffuser flow is demonstrated using the pressure coefficient, $C_p = (p - p_{ref}) / (0.5 \rho U_b^2)$ along the streamwise coordinate $\pos[1]/L$, cf.~\cref{fig:diffuser_incomp_cp_umean} (left).
Following~\citet{cherry2008geometric}, the reference pressure $p_{ref}$ is taken at $\pos[1]/L = 0.045$.
The experimental data of~\citet{cherry2008geometric} and the numerical results of~\citet{ohlsson2010direct} and~\citet{miro2024self} serve as a reference.
The simulation in~\cite{ohlsson2010direct} utilized a high-order \nth{12} order spectral element method, while the simulation
of~\citet{miro2024self} was performed using a second-order Galerkin finite element method.
Convergence of the LES results to the experimental reference is demonstrated in~\cref{fig:diffuser_incomp_cp_umean} (left) for the pressure coefficient by using LES-G with an increased resolution ($\ppn=7$), and a further increase to $\ppn=11$ has no significant impact.
A direct comparison of the LES-G with the LES-GL simulations reveals that the former is in better agreement with the
experimental reference for a fixed polynomial degree.
In addition, the LES-GL simulation at a higher polynomial degree of $\ppn=7$ is not yet converged to the reference.
Another important finding is that a polynomial degree of $\ppn=7$ has to be used for Gauss--Lobatto nodes in order to achieve approximately the same accuracy as for the calculation with Gauss--Legendre nodes at $\ppn=5$.
This, in turn, results in an even smaller time step compared to Gauss--Legendre nodes at $\ppn=5$.
Consequently, the LES with Gauss--Legendre nodes remains the more efficient option in terms of the achieved accuracy for the given computational resources.

A similar behavior is visible for the mean center plane velocity profile, cf.~\cref{fig:diffuser_incomp_cp_umean} (right).
The results demonstrate a better agreement of the LES-G simulation with the log law and the numerical reference. Conversely, the LES-GL simulation overestimates the log law region.
As illustrated in~\cref{fig:diffuser_incomp_cp_umean} (right), the results for the mean center plane velocity profile
demonstrate that the LES-G simulation has not yet fully converged, in contrast to the pressure distribution.
In view of the fact that the objective of the present study is to conduct an LES rather than a direct numerical simulation for which the difference between Gauss--Legendre and Gauss--Lobatto nodes would vanish, the LES results are considered validated.

One of the key parameters for the evaluation of diffuser's is the pressure recovery over the diffuser section, $\Delta C_p = C_p(\pos[1]/L = 1) - C_p(\pos[1]/L = 0.045)$, where even a slight error reduction is of paramount importance.
The findings in~\cref{tab:diffusor_incomp_pressure_recovery} demonstrate that both LES-GL and LES-G simulations slightly
overestimate the pressure recovery.
However, it is evident that the entropy stable DGSEM on Gauss–Legendre nodes achieves superior results by predicting the pressure
recovery more accurately for the given grid resolution, leading to a significant error reduction of 37\% compared to the
experimental reference.

\begin{table}[tbp]
  \centering
  \resizebox{\textwidth}{!}{\begin{tabular}{ccccccc}
                    & Cherry et al. & Miró et al. & LES-GL  & LES-GL ($\ppn=7$)  & LES-G   & LES-G ($\ppn=11$) \\ \hline
      $\Delta C_p$  & $0.527$       & $0.500$     & $0.602$ & $0.560$            & $0.574$ & $0.542$ \\ \hline
  \end{tabular}}
  \caption{Pressure recovery over the diffuser section for the LES-GL and LES-G simulations using $\ppn=5$ and results with increased resolution. The experimental data of~\citet{cherry2008geometric} and the numerical results of~\citet{miro2024self} serve as a reference.}
  \label{tab:diffusor_incomp_pressure_recovery}
\end{table}

\begin{figure}[tbp]%
  \centering
  \includestandalone[width=0.8\linewidth]{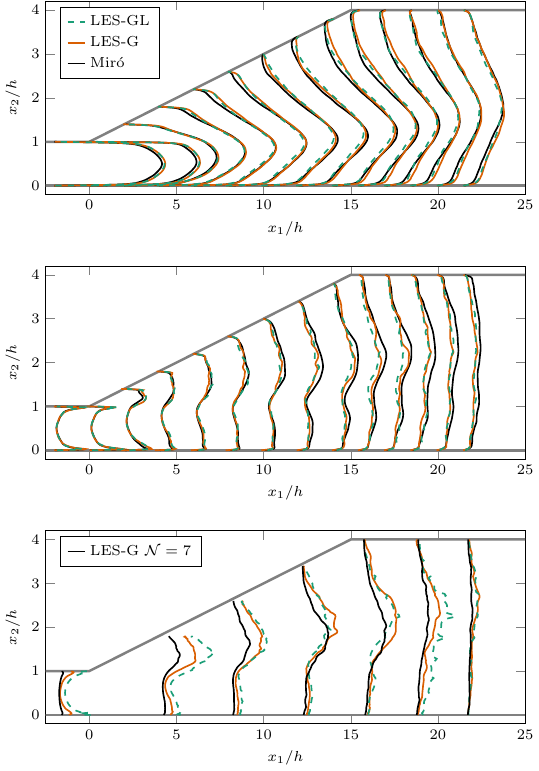}
  \caption{
    Comparison of the LES results using Gauss--Legendre (LES-G) and Gauss--Lobatto (LES-GL) nodes to simulation results of~\citet{miro2024self} for the diffuser inflow bulk Mach number of $\MaIn=0.2$.
    Top: Mean center plane velocity at $\pos[3]/b=0.5$, scaled as $5 \overline{U} + \pos[1]/L$.
    Mid: Normal Reynolds stress $\overline{u'u'}$ at $\pos[3]/b=0.5$, given as $50 \overline{u'u'} + \pos[1]/L$.
    Bottom: Pressure fluctuation $\overline{p'p'}$ at $\pos[3]/b=0.5$ given as $2000 \overline{p'p'} + \pos[1]/L$ at $\pos[1]/L \in
  \{-2;4;8;12;15.5;18.5;21.5 \}$. With no literature data available, the results are compared to a better-resolved LES (LES-G $\mathcal{N}=7$).
The pressure fluctuations are non-zero at the wall due to the absence of a no-slip condition.}
  \label{fig:diffuser_UMean_Ma0p2_1}
\end{figure}

A more quantitative comparison of the wall-normal mean velocity $\overline{U}$ profiles in the spanwise center plane is illustrated in~\cref{fig:diffuser_UMean_Ma0p2_1} (top) for a bulk inflow Mach number of $\MaIn =0.2$.
Here, $\overline{(\cdot)}$ denotes a time-averaged quantity.
The numerical results of~\citet{miro2024self} serve as a reference.
The findings indicate that both LES-GL and LES-G simulations demonstrate the ability to adequately capture the flow
field.
Slight differences to the reference are visible at the diffuser inflow. This phenomenon can be attributed to the influence of weak compressibility in the $M_{b,in}=0.2$ regime, where the density falls below the initial value of 1, thereby increasing the velocity. It is important to note that there is a strong correlation between the velocity results and the numerical reference.
It is evident that there is an absence of a discernible trend with regard to the performance of the LES-G and LES-GL simulation in the separated flow region and in the duct section following the diffusers.
A similar behavior can be observed for the normal component of the Reynolds stresses $\overline{u'u'}$, cf.~\cref{fig:diffuser_UMean_Ma0p2_1} (mid).
As demonstrated in~\cref{fig:diffuser_UMean_Ma0p2_1}, an increase in the diffuser cross-section is observed to correspond with
an increase in deviations from the reference mean velocity field and normal Reynolds stress.
These deviations can be attributed to the decreasing effective resolution along the diffuser.
Consequently, the error resulting from the mesh resolution becomes predominant in comparison to the discrepancies between LES-G and LES-GL.
A similar trend is also observed for the pressure fluctuations illustrated in~\cref{fig:diffuser_UMean_Ma0p2_1} (bottom).
In addition, compared to the LES-GL results, the solution of LES-G is found to be much smoother. Conversely, the LES-GL results
reveal the presence of unphysical kinks in the pressure fluctuations which correspond to the element boundaries.
This behaviour is not generally considered to be desirable, as it is not comprehensible from a physical point of view.
Consequently, the Gauss--Legendre nodes manifest superior characteristics.
Here, only the normal Reynolds stress and pressure fluctuations are considered as the remaining Reynolds stresses and higher statistical moments demonstrate a comparable behaviour.

To summarize, the results illustrate that both solutions are generally able to reproduce the reference well, but a discernible trend concerning the performance of the LES-G and LES-GL simulations remains elusive.
This is to be expected since integral quantities such as the pressure coefficient converge faster compared to first or high-order statistical moments. Consequently, less differences will be visible for the latter.
It can thus be concluded that given the significantly superior prediction of $C_p$ and the mean velocity field, LES-G can be
considered as the more accurate solution.

Finally, it is important to note that the enhanced accuracy of Gauss--Legendre nodes comes at the expense of a smaller time step and an
increase in computing time (due to the entropy projection and additional surface fluxes, cf.~\cref{sec:methods}).
Nevertheless, it can be shown that in order to achieve an accuracy comparable to that of the Gauss--Legendre nodes at $\ppn = 5$, a polynomial degree of $\ppn = 7$ is necessary for Gauss--Lobatto nodes cf.~\cref{fig:diffuser_incomp_cp_umean} (left).
It is evident that this results in a time step which is even smaller than that of Gauss--Legendre nodes at $\ppn = 5$.
Consequently, the LES-G simulation is more efficient in terms of the accuracy gained per computational resource expended, as compared to LES-GL.

\subsubsection{Compressible diffuser flow}
\begin{figure}[tbp]%
  \centering
  \includestandalone[width=\linewidth]{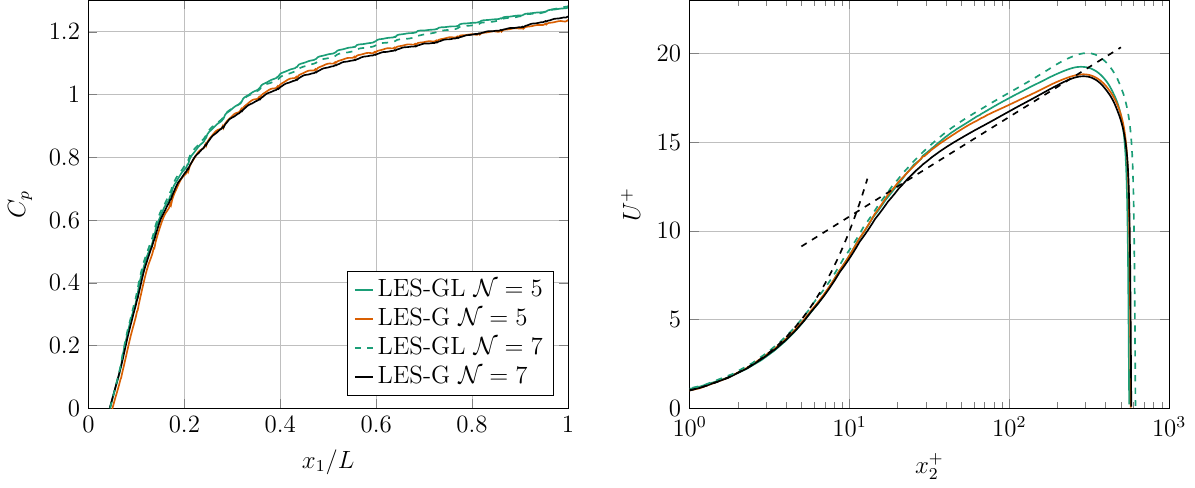}
  \caption{Comparison of the current LES results using Gauss--Legendre (LES-G) and Gauss--Lobatto (LES-GL) nodes for a diffuser
    inflow bulk Mach number of $\MaDIn=0.7$.
    A better resolved LES at $\ppn =7$ using Gauss--Legendre nodes (LES-G $\ppn =7$) serves as a reference.
    Left: Pressure coefficient on the bottom wall $\pos[2]/h=0$ at $\pos[3]/b=0.5$.
    Right: Mean center plane velocity profile at $2h$ upstream of the diffuser inflow.
    The log law $\log(\pos[2]^+)/\kappa+B$ (dashed line) is calculated using $\kappa = 0.41$ and $B=5.2$.
  }
  \label{fig:diffuser_cp_2}
\end{figure}

Increasing the Mach number at the diffuser inflow to the compressible regime with a value of $\MaDIn =0.7$ results in more pronounced differences between the LES-GL and LES-G simulations.
The absence of literature data precludes an exact judgment; however, with the weakly compressible case validated and the
observations for the compressible TGV in mind, cf.~\cref{sec:tgv_comp}, the following results can be considered sufficiently
accurate to allow for a comparison of both simulations.
In order to provide a numerical reference for this case and thereby facilitate the identification of trends, the LES-G simulation was carried out with an increased polynomial degree of $\ppn =7$.

Starting with the pressure distribution, the results shown in~\cref{fig:diffuser_cp_2} (left) clearly demonstrate a similar behavior to that observed in the previous test cases: The LES-G simulation is more accurate compared to LES-GL for a fixed polynomial degree.
In addition, the LES-G simulation at $\ppn=5$ is already converged for the pressure distribution, and an increase to $\ppn=7$ has no significant impact.
Conversely, convergence has not yet been observed for the LES-GL results, even for a LES with higher resolution at $\ppn=7$.
In addition, in contrast to the weakly compressible case, the LES-G simulation at $\ppn=5$ shows superior accuracy in comparison to LES-GL.
A similar behavior is visible for the mean center line velocity profile, cf.~\cref{fig:diffuser_cp_2} (right).
Analogous to the weakly compressible case, the mean center line velocity profile requires an even finer-resolved LES to demonstrate convergence. However, this is not the focus of the present work.

Analyzing again the pressure recovery, $\Delta C_p = C_p(\pos[1]/L = 1) - C_p(\pos[1]/L = 0.045)$, values of $\Delta C_{p,GL} =
1.275$ and $\Delta C_{p,G} = 1.263$ are obtained for the LES-GL and LES-G simulations with $\ppn=5$, respectively.
Again, the results obtained using the Gauss--Legendre nodes demonstrate a higher agreement with the pressure recovery of
$\Delta C_{p,ref} = 1.242$ predicted by the reference LES with $\ppn=7$.

\begin{figure}[tbp]%
  \centering
  \includestandalone[width=0.8\linewidth]{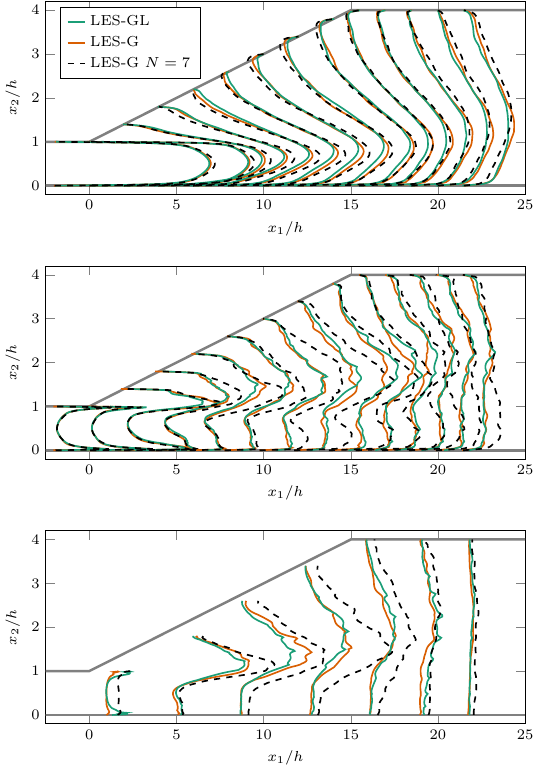}
  \caption{Comparison of the LES results using Gauss--Legendre (LES-G) and
Gauss--Lobatto (LES-GL) nodes for a diffuser inflow bulk Mach number of $\MaDIn=0.7$.
  A better resolved LES at $\ppn =7$ using Gauss--Legendre nodes (LES-G $\ppn =7$) serves as a reference.
  Top: Mean center plane velocity at $\pos[3]/b=0.5$, given as $5 \overline{U} + \pos[1]/L$. Mid: Normal Reynolds stress
  $\overline{u'u'}$ at $\pos[3]/b=0.5$ given as $50 \overline{u'u'} + \pos[1]/L$.
Bottom: Pressure fluctuation $\overline{p'p'}$ at $\pos[3]/b=0.5$ given as $2000 \overline{p'p'} + \pos[1]/L$ at $\pos[1]/L \in \{0;4;8;12;15.5;18.5;21.5\}$.
The pressure fluctuations are non-zero at the wall due to the absence of a no-slip condition.}
  \label{fig:diffuser_UMean_Ma0p45_1}
\end{figure}

The pronounced differences between the LES-GL and LES-G simulations can also be observed in the wall-normal mean velocity profiles in the spanwise center plane illustrated in~\cref{fig:diffuser_UMean_Ma0p45_1} (top).
The results clearly show that the LES-G simulations are in better agreement with the better-resolved reference, especially in the front section of the diffuser.
This phenomenon can be attributed to the enhanced resolution, due to the grid stretching along the diffuser in the front section of the diffuser.

A similar behavior can also be seen in the normal Reynolds stress $\overline{u'u'}$ depicted in~\cref{fig:diffuser_UMean_Ma0p45_1} (mid), particularly in the well-resolved section directly within the diffuser, where the formation of the shear layer becomes apparent.
The results demonstrate that the LES-G simulation is able to match the numerical reference better as compared to LES-GL. This trend persists along the diffuser path, albeit with a gradual decline in intensity.
A similar trend is also observed for the pressure fluctuations illustrated in~\cref{fig:diffuser_UMean_Ma0p45_1} (bottom).
Furthermore, the unphysical oscillations mentioned in~\cref{sec:weaklycomprdiff} are once again evident.
As above, only the normal Reynolds stress and the pressure fluctuations have been considered since the remaining Reynolds stresses and high-order statistical moments demonstrate a comparable behaviour.

In summary, the Gauss--Legendre nodes demonstrate slightly superior accuracy in comparison to the Gauss--Lobatto nodes. This is particularly evident in the pressure distribution shown for the weakly compressible case, but also in the central characteristics such as the precise localization of the velocity maximum within the diffuser.
This observation is also confirmed when the Mach number is increased, where even greater discrepancies in the results are
evident, particularly in the higher statistical moments. A comparison with a higher-resolution numerical simulation once again demonstrates the enhanced accuracy of the Gauss–Legendre nodes.
More detailed investigations are beyond the scope of this work and will be detailed in further in-depth studies.

\subsection{Computational cost}
Finally it has to be noted that the higher accuracy of Gauss--Legendre nodes comes at the expense of longer computing time,
approximately by a factor of 2 for all considered test cases and a fixed polynomial degree.
For the compressible case, the LES-GL requires \num{27890} CPU-h, whereas the LES-G necessitates \num{62080} CPU-h.
The LES of the weakly compressible case requires nearly twice as much computing time as its compressible counterpart since the
higher Mach number allows overall greater time steps. %
The reason for this is that the use of Gauss–Lobatto nodes allows a larger time step, whereas the time step of Gauss–Legendre nodes is more restrictive.
The elevated computational costs are attributable to the entropy projection and additional surface fluxes required for ESDGSEM-G.
However, as discussed in \cref{sec:weaklycomprdiff}, despite the higher computational costs, it is more efficient to achieve a comparable result with Gauss--Legendre nodes.
The enhanced accuracy of Gauss--Legendre nodes necessitates a lower resolution, consequently resulting in elevated overall computational costs when employing Gauss--Lobatto nodes.

\section{Conclusion}
\label{sec:conclusion}

High-order methods rely on adequate stabilization techniques for the simulation of compressible, turbulent flows.
The key contribution of this work is a detailed comparison of the performance of entropy stable DGSE schemes on Gauss–Legendre and Gauss–Lobatto nodes. This investigation encompasses complex, compressible turbulent flows, with a particular emphasis on the evaluation of accuracy, stability, and efficiency.
The results have shown that the DGSEM on Gauss--Lobatto nodes is less accurate compared to Gauss--Legendre nodes due to the lower integration accuracy, as anticipated.
This difference is especially pronounced for more compressible turbulent flows, particularly with regard to the higher statistical moments such as Reynolds stresses or pressure fluctuations.
Conversely, simulations with DGSEM on Gauss--Legendre nodes are more expensive, owing to the more restrictive time step and higher computational load involved.
The elevated computational load is attributable to the entropy projection and additional surface fluxes.
Nevertheless, it has been demonstrated that in order to attain a comparable level of accuracy with Gauss--Lobatto nodes as that
observed with Gauss--Legendre nodes, a higher resolution is necessary for the former, leading to a more constrained time step.
It can thus be concluded that the LES on Gauss–Legendre nodes is the more efficient option, achieving a comparable error to a more refined and consequently more expensive LES on Gauss–Lobatto nodes.

Finally, it is imperative to acknowledge that the findings reported herein are contingent upon the considered applications.
Further detailed studies are requisite to broaden the scope of the present findings, including more complex flows such as a shock
turbulent boundary layer interaction.

\section*{Acknowledgements}
This work was funded by the European Union and has received funding from the European High Performance Computing Joint Undertaking (JU) and Sweden, Germany, Spain, Greece, and Denmark under grant agreement No 101093393.
The research presented in this paper was funded in parts by Deutsche Forschungsgemeinschaft (DFG, German Research
Foundation) under Germany's Excellence Strategy - EXC 2075 - 390740016 and by the state of Baden-Württemberg under the project Aerospace 2050 MWK32-7531-49/13/7 "QUASAR".
We acknowledge the support by the Stuttgart Center for Simulation Science (SimTech).
Further, we want to gratefully acknowledge funding by the DFG through SPP 2410 Hyperbolic Balance Laws in Fluid Mechanics: Complexity, Scales, Randomness (CoScaRa).
The authors gratefully acknowledge the support and the computing time on ``Hawk'' provided by the HLRS through the project ``hpcdg''.

\noindent \section*{In memoriam}
\noindent This paper is dedicated to the memory of Prof. Arturo Hidalgo L\'opez
($^*$July 03\textsuperscript{rd} 1966 - $\dagger$August 26\textsuperscript{th} 2024) of the Universidad Politecnica de Madrid,
organizer of HONOM 2019 and active participant in many other editions of HONOM.
Our thoughts and wishes go to his wife Lourdes and his sister Mar\'ia Jes\'us, whom he left behind.

\bibliographystyle{elsarticle-num-names}
\bibliography{references}

\end{document}